\documentclass[twocolumn,prb,aps,superscriptaddress]{revtex4-1}
\usepackage{graphicx}
\usepackage{siunitx}
\usepackage{hyperref}

\begin{document}
\title{Tuning non-collinear magnetic states by hydrogenation}
\author{Aurore Finco}
\email{aurore.finco@normalesup.org}
\altaffiliation[Present address: ]{Laboratoire Charles Coulomb, Université de Montpellier and CNRS, F-34095 Montpellier, France.}
\author{Pin-Jui Hsu}
\altaffiliation[Present address: ]{Department of Physics, National Tsing Hua University, 30013 Hsinchu, Taiwan.}
\author{Kirsten von Bergmann}
\author{Roland Wiesendanger}
\affiliation{Department of Physics, University of Hamburg, D-20355 Hamburg, Germany}

\date{\today}

\begin{abstract}
  Two different superstructures form when atomic H is incorporated in the Fe monolayer on Ir(111). Depending on the amount of H provided, either a highly ordered p$(2\times2)$ hexagonal superstructure or an irregular roughly square structure is created. We present here spin-polarized scanning tunneling microscopy (SP-STM) measurements which reveal that in both cases the magnetic nanoskyrmion lattice state of the pristine Fe monolayer is modified. Our measurements of the magnetic states in these hydrogenated films are in agreement with superpositions of cycloidal spin spirals which follow the pattern and the symmetry dictated by the H superstructures. We thus demonstrate here the possibility to vary the symmetry of a non-collinear magnetic state in an ultrathin film without changing its substrate. 
\end{abstract}

\maketitle

\section{Introduction}
A crucial issue for the  stabilization of non-collinear magnetic states in ultrathin films is the ability to tune their properties. To achieve this goal, several ways are explored, mostly relying on the choice of appropriate materials and interface engineering~\cite{fert_skyrmions_2013, dupe_engineering_2016, buttner_theory_2018}. These approaches have been successfull in the last years, allowing to stabilize magnetic skyrmions at room temperature~\cite{jiang_blowing_2015, moreau-luchaire_additive_2016, boulle_room-temperature_2016} and to explore their dynamical properties owing to the improvement of structural properties of the magnetic films~\cite{woo_observation_2016, jiang_direct_2017, litzius_skyrmion_2017}.  Recently, another option to tune non-collinear magnetism was proposed, the incorporation of atomic H in an ultrathin magnetic film. In particular, this was realized in the Fe double layer on Ir(111)~\cite{hsu_inducing_2018}. The magnetic state of the pristine Fe double layer is a cycloidal spin spiral with a short period of about \SI{1.5}{\nano\meter}~\cite{hsu_guiding_2016}. No change of the magnetic configuration has been observed when an out-of-plane magnetic field up to \SI{9}{\tesla} is applied, which means that a skyrmionic phase does not appear. Upon hydrogenation, two different superstructures form in the Fe double layer~\cite{hsu_inducing_2018}: a p$(2\times2)$ superstructure (H1 phase) with a structural period of \SI{0.54}{\nano\meter} and a larger hexagonal superstructure (H2 phase) with a period of about \SI{1}{\nano\meter}. The H2 phase is ferromagnetic and the H1 phase exhibits a spin spiral state with a magnetic period of about \SI{3.5}{\nano\meter}. In addition to this nearly three times larger magnetic period compared to the pristine Fe double layer, skyrmions appear in the H1 phase under an out-of-plane magnetic field of \SI{3}{\tesla}, demonstrating that hydrogenation is a new tool which can be used for the engineering of complex magnetic states.

In the present article, we investigate the structure and the magnetic state of the hydrogenated Fe monolayer on Ir(111) at the nanoscale using spin-polarized scanning tunneling microscopy (SP-STM)~\cite{wiesendanger_spin_2009}. Previous detailed studies have shown that the pristine Fe monolayer on Ir(111) is characterized by a complex non-collinear magnetic state called a nanoskyrmion lattice~\cite{heinze_spontaneous_2011}. This magnetic structure is stabilized by the competition between the different magnetic energy contributions in the system. Besides the exchange coupling, the magnetocrystalline anisotropy and the Dzyaloshinkii-Moriya interaction (DMI), it is necessary to take into account higher-order terms (four-spin and biquadratic interactions) in order to understand the stabilization of this nanoskyrmion lattice~\cite{heinze_spontaneous_2011}. Depending on the stacking of the atomic Fe monolayer, the characteristics of the nanoskyrmion lattice are different. In the fcc Fe monolayer, the nanoskyrmion lattice is square and incommensurate with the atomic lattice~\cite{heinze_spontaneous_2011, grenz_probing_2017, hauptmann_sensing_2017}, whereas in the hcp Fe layer, the nanoskyrmion lattice is hexagonal and commensurate with the atomic structure~\cite{von_bergmann_influence_2015}. In both cases, the magnetic period is about \SI{1}{\nano\meter}.

Here we report the formation of two different superstructures in the Fe monolayer by the incorporation of different amounts of atomic H. In both cases, the nanoskyrmion lattice is changed by the presence of the H superstructure, which illustrates the possibility to modify complex non-collinear magnetic states using atomic H.   

\section{Methods}

The sample preparation and the STM measurements were performed in an ultrahigh vacuum system consisting of several chambers dedicated either to a particular step of the preparation or to the measurements. The base pressure of the vacuum system is about \SI{1e-10}{\milli\bar}.
The (111) surface of the Ir single crystal was cleaned by repeated cycles of Ar-ion sputtering at \SI{800}{\electronvolt} and annealing up to \SI{1200}{\celsius} for \SI{90}{\second}. In addition, oxygen annealing of the Ir(111) substrate was performed by varying the substrate temperature from room temperature to \SI{1350}{\celsius} under an oxygen pressure between \SI{5e-6}{\milli\bar} and \SI{5e-8}{\milli\bar}. This procedure removes the C contamination at the surface.

Between 1 and 1.5 atomic layers (AL) of Fe is deposited onto the clean substrate at elevated temperature, shortly after the annealing, at a rate of about 1 AL per min. The Fe monolayer grown using this procedure exhibits an fcc stacking~\cite{heinze_spontaneous_2011}. After the Fe deposition, we wait for the sample to cool down to room temperature before exposing it to atomic H. The atomic H is produced by cracking H$_2$ molecules at high temperature in a dedicated source. The amount of H atoms required to create the superstructures depends strongly on the Fe coverage, we use a partial pressure of \SI{1e-8}{\milli\bar} for an exposure time between 2 and \SI{10}{\minute}. After the H exposure, the sample is post-annealed at \SI{300}{\celsius} for \SI{10}{\minute} to favor the formation of the superstructures and to remove the excess H which would make the STM measurements more difficult. This cycle of H exposure/post-annealing can be repeated in order to increase the amount of H atoms incorporated in the Fe film.

The measurements were performed in two different home-built low-temperature STM setups operating at \SI{4}{\kelvin} or \SI{8}{\kelvin}, respectively. We used either chemically etched antiferromagnetic Cr bulk tips or ferromagnetic Fe-coated W tips. We use the known spin spiral state present in the hydrogenated Fe double layer H1 areas~\cite{hsu_inducing_2018} on the sample to determine the magnetization axis of the tip.

The SP-STM simulations are performed using the procedure described in ref.~\cite{heinze_simulation_2006}. The magnetic contrast is assumed to originate only from the tunneling magnetoresistance.

\section{Results}

\subsection{Structure}

\begin{figure}
  \centering
  \includegraphics{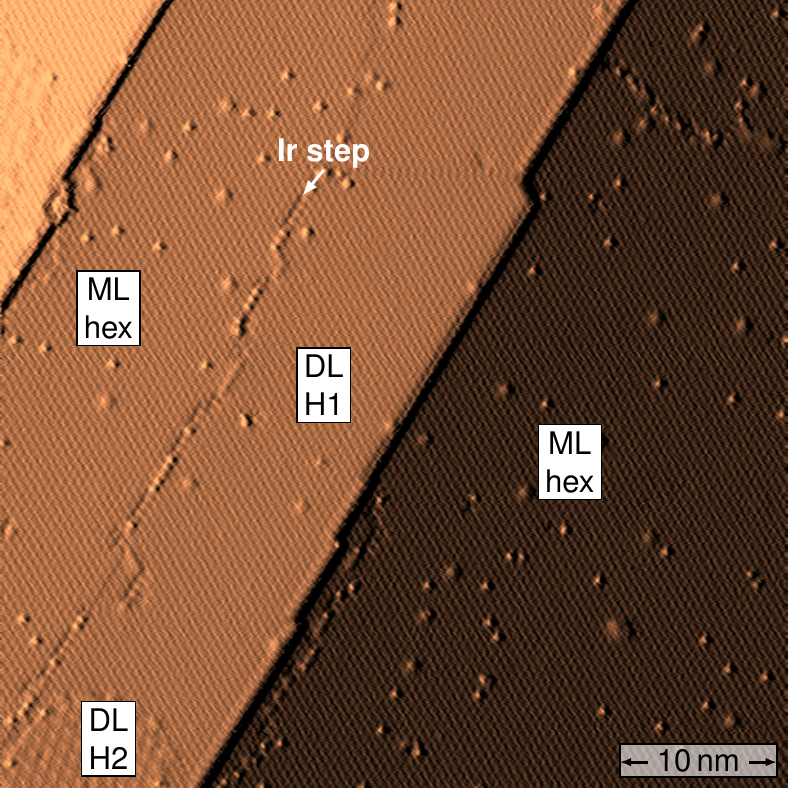}
  \caption{\textbf{Overview of a hydrogenated sample.} STM constant-current map of a hydrogenated Fe film on Ir(111). Hexagonal superstructures cover completely the monolayer (ML) and the double layer (DL) areas. In the latter case, both the H1 and the H2 phases are visible. The image is partially differentiated along the scan direction (horizontal) to improve the visibility of the structures.
    Measurement parameters:~\SI{700}{\milli\volt}, \SI{1}{\nano\ampere}, \SI{8}{\kelvin}, \SI{0}{\tesla}.}
  \label{fig:overview}
\end{figure}

\begin{figure}
  \centering
  \includegraphics{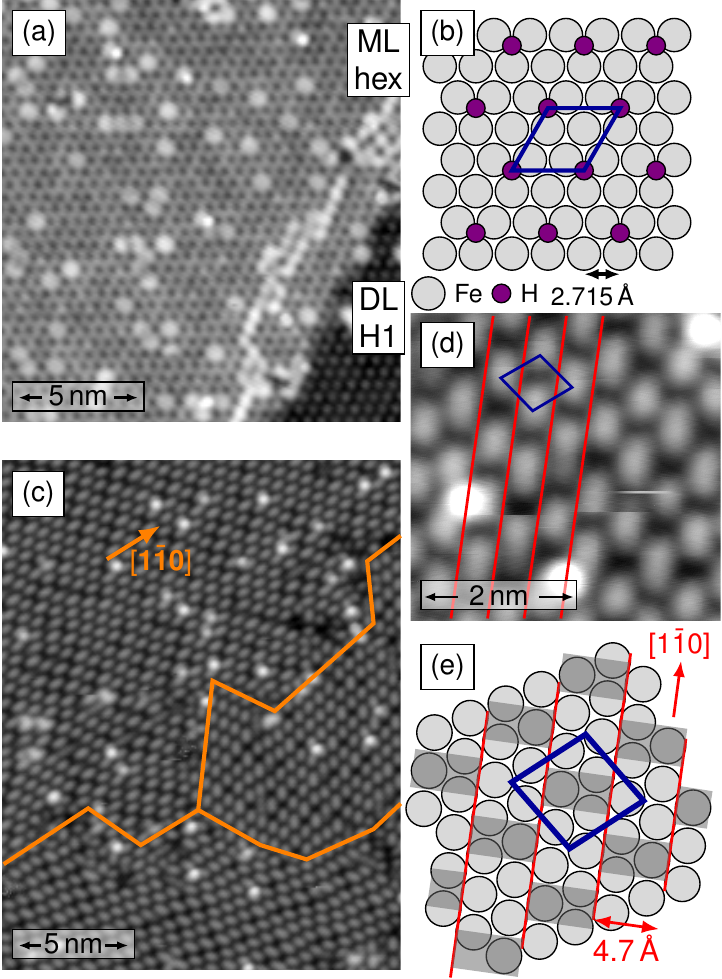}
  \caption{\textbf{The two phases of the hydrogenated Fe monolayer (ML) on Ir(111).} (a) STM constant-current map showing the H-induced hexagonal superstructure. The darker area corresponds to the H1 phase of the hydrogenated Fe double layer (DL) located on the adjacent terrace, which allows to identify the hexagonal structure in the monolayer as a p$(2\times2)$ superstructure. (b) Proposed lateral position of the H atoms in the hexagonal superstructure. The blue line marks the unit cell. (c) STM constant-current map showing the three rotational domains of the square superstructure, the domain boundaries are marked with the orange lines. (d) STM constant-current map showing a closer view of the square superstructure. The bright regions are aligned along the close-packed rows of the (111) surface and form an irregular roughly square pattern. The blue line marks the structural unit cell. (e) Sketch showing the position of the square superstructure on the atomic lattice. The dark areas in image (d) are marked in dark gray. A row of bright dots corresponds to 2 rows of atoms. The unit cell of the superstructure is indicated in blue.
    Measurement parameters:~(a):~\SI{50}{\milli\volt}, \SI{1}{\nano\ampere}, \SI{8}{\kelvin}; (c):~\SI{800}{\milli\volt}, \SI{1}{\nano\ampere}, \SI{4}{\kelvin}; (d):~\SI{500}{\milli\volt}, \SI{1}{\nano\ampere}, \SI{4}{\kelvin}.}
  \label{fig:structure}
\end{figure}

Figure~\ref{fig:overview} shows an overview of a sample prepared by the incorporating atomic H in an Fe ultrathin film on Ir(111) measured with STM. The Fe coverage here is around 1.3 AL and we see monolayer and double layer areas. The H atoms are first incorporated in the Fe double layer regions. If the provided H amount is sufficient to cover fully the Fe double layer areas with the H1 and H2 superstructures as it is the case here, we observe the incorporation of H atoms in the Fe monolayer. If only a small amount of H atoms is available for incorporation in the Fe monolayer, small patches of a hexagonal structure form. The area occupied by these patches increases with the amount of H provided, until the Fe monolayer is fully covered by the hexagonal superstructure as in figure~\ref{fig:overview}. Since the Fe film is deposited at elevated temperature, the stacking of the pristine Fe layer is fcc~\cite{heinze_spontaneous_2011}. We thus also expect that the Fe atoms in the hydrogenated Fe monolayer are located in the fcc sites. A more detailed view of the hexagonal superstructure is presented in figure~\ref{fig:structure}(a). A buried Ir step edge is present at the bottom right corner, showing the boundary between the hexagonal phase in the hydrogenated Fe monolayer and the H1 phase in the hydrogenated Fe double layer (bottom right corner, darker area). Comparison of the two patterns shows that they have the same size and the same orientation, which means that they are both p$(2\times2)$ superstructures~\cite{hsu_inducing_2018}. The boundary between the two areas is very smooth, suggesting that the transition between the two atomic arrangements is continuous. Since the bright protrusions in the Fe double layer correspond to H vacancies, we assume that the dark dots in the Fe monolayer mark the positions of the H atoms and derive the structure model shown in Figure~\ref{fig:structure}(b). According to this model, the H concentration in the monolayer is half of that of the H1 phase of the double layer. Note that one cannot know from the STM data if the H atoms are located on top of the Fe layer or at the Fe/Ir interface.

A second H superstructure can be prepared in the Fe monolayer when a sample exhibiting this p$(2\,\times~2)$ superstructure is exposed to atomic H at room temperature and post-annealed again. Small patches of this new superstructure form in the hexagonal phase and if enough H atoms are provided, it can cover the complete Fe monolayer as illustrated in Figure~\ref{fig:structure}(c). We did not succeed in preparing this new superstructure using only one cycle of H exposure/post-annealing. It seems that it is easier to create this structure once the hexagonal superstructure is formed rather than starting directly from the pristine Fe film. Interestingly, the symmetry of this superstructure is different from the previously observed hexagonal phases. Here, the superstructure is roughly square. Consequently, since the atoms of the Ir(111) surface form a hexagonal lattice, three rotational domains of this superstructure can be found (separated by the orange lines in Figure~\ref{fig:structure}(c)). In each domain, one direction of the diagonals of the squares is oriented along a close-packed row of the (111) surface. In addition, this superstructure is not fully regular. Besides the presence of defects, the size of the bright regions is varying and their arrangement is not fully periodic. A closer view of the structure is presented in Figure~\ref{fig:structure}(d), showing that it is commensurate with the substrate along the $[11\bar{2}]$ direction but varying along the $[1\bar{1}0]$ direction. Each row of bright dots corresponds to two rows of Fe atoms as depicted in the sketch in Figure~\ref{fig:structure}(e). The bright regions are elongated along this direction and their length and spacing is not fixed. An approximated unit cell for the structure is indicated in blue in the figures~\ref{fig:structure}(d) and (e). The H concentration is expected to be larger in this roughly square superstructure than in the p$(2~\times~2)$ superstructure but we cannot determine the exact position of the H atoms from the STM data. As indicated in the sketch in figure~\ref{fig:structure}(e), it seems that more H atoms are located in the areas marked with the dark gray rectangles than the others. The square superstructure is very mobile, some jumps of the bright dots are visible in the scan shown in figure~\ref{fig:structure}(d) as straight horizontal lines. These movements might be induced by the STM tip and their presence suggests that some of the H atoms could be adsorbed at the surface of the Fe film rather than at the Fe/Ir interface.

\subsection{Magnetism}

\begin{figure}
  \centering
  \includegraphics{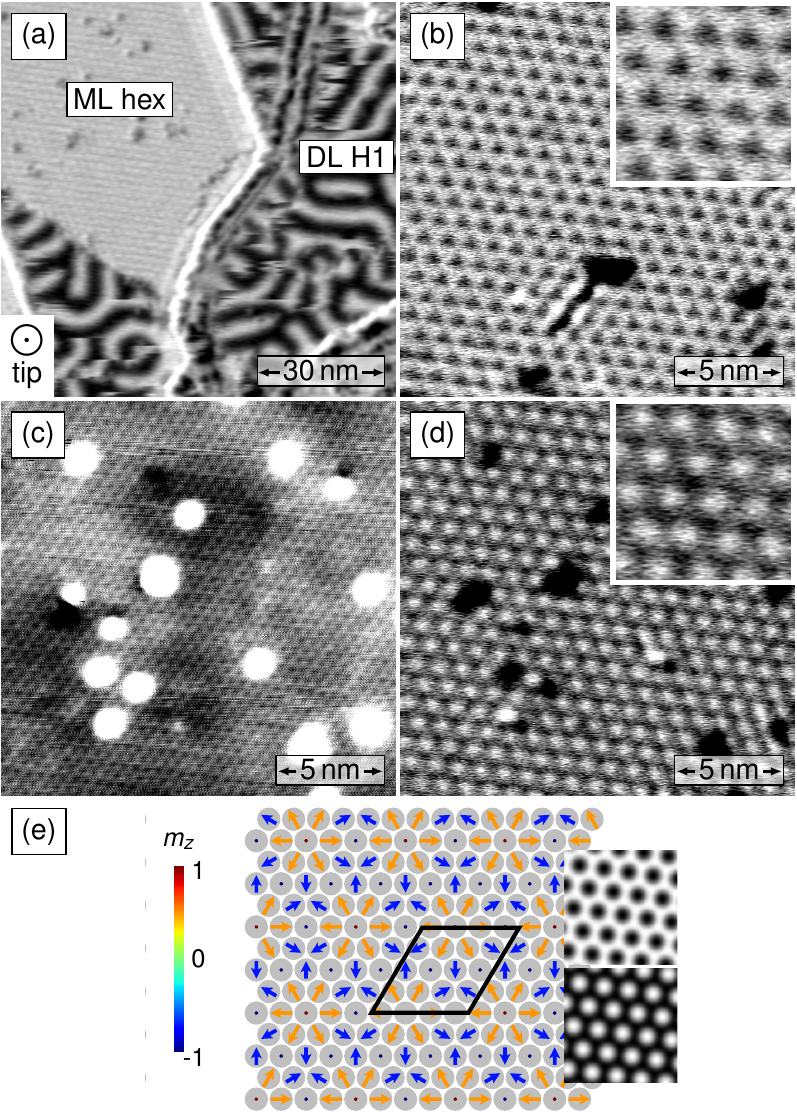}
  \caption{\textbf{Magnetic state of the hexagonal phase of the hydrogenated Fe monolayer on Ir(111).} (a) Spin-resolved differential conductance map showing an area of the hydrogenated Fe monolayer (ML) and the spin spiral in the neighboring H1 region of the double layer (DL). The spin spiral is used as a reference to determine the magnetization axis of the tip (here out-of-plane). (b) and (d) Differential conductance maps of two different sample areas showing the magnetic state of the hexagonal phase, measured with the same out-of-plane sensitive tip as in (a). The areas displayed in (b) and (d) correspond to opposite magnetic domains. The insets show magnified views of the data. (c) STM constant-current map of the region shown in (d). No magnetic contrast is visible, which allows to see the H superstructure and to conclude that the magnetic unit cell corresponds to a p$(4\times4)$ structure. Measurement parameters:~\SI{-1}{\volt}, \SI{1}{\nano\ampere}, \SI{4}{\kelvin}, Cr bulk tip. (e) Representation of a possible hexagonal nanoskyrmion lattice state. In this configuration, the points where the out-of-plane component of the magnetization is maximal are located on top sites of the atomic lattice (on-top state). The insets show SP-STM simulation of the two magnetic domains with an out-of-plane sensitive magnetic tip, displayed with the same lateral scale as the data in (b)-(d). The SP-STM simulations were performed with the following realistic parameters: tip-sample distance:~\SI{600}{\pico\meter}, work function:~\SI{4.8}{\electronvolt}, spin polarization of sample (s) and tip (t):~$P_sP_t=0.2$.}
  \label{fig:mag_hex}
\end{figure}

The magnetic state of the hexagonal phase of the hydrogenated Fe monolayer is shown in figure~\ref{fig:mag_hex}. The images (a), (b) and (d) are differential conductance maps measured with the same Cr bulk tip. From the magnetic contrast observed on the spin spiral of the H1 region of the double layer in the image~\ref{fig:mag_hex}(a), we derive that the magnetization axis of the tip is out-of-plane. The images (b) and (d) show two different areas exhibiting an hexagonal magnetic pattern. In the region (b), the pattern appears as dark dots on a bright background whereas in the region (d), it constits of bright dots on a dark background. These two areas correspond to opposite magnetic domains. The two insets show a magnified view of the data. The STM constant-current map recorded simultaneously with image (d) is presented in (c). Since no magnetic signal is visible in image (c), the H-induced structural unit cell dominates. It follows from the comparison between (c) and (d) that the magnetic pattern corresponds to a p$(4\times4)$ superstructure.

The magnetic domains can be switched when an out-of-plane magnetic field of $\SI{\pm 9}{\tesla}$ is applied. This confirms that the observed domains have opposite magnetizations and that the magnetic unit cell possesses an uncompensated out-of-plane magnetic moment. This state is reminiscent of the hexagonal nanoskyrmion lattice present in the pristine hcp Fe monolayer on Ir(111)~\cite{von_bergmann_influence_2015} but with a larger magnetic unit cell here. These observations suggest that the magnetic state of this hydrogenated hexagonal phase is a hexagonal nanoskyrmion lattice which can be seen as the superposition of three symmetry-equivalent cycloidal spin spirals. There are several possibilities to choose the positions of the three spin spirals with respect to each other and the atomic lattice. However, from the symmetry and the period of the measured pattern, only two options are possible.  Either the point of the magnetic structure where the out-of-plane magnetic moment is maximum corresponds to a top site of the atomic lattice (on-top state) or to a hollow site (hollow state). To obey the constraint of constant magnetic moment, the size of the magnetic moments is normalized after the superposition. Since these states cannot be distinguished from our experimental data, only the on-top state is sketched in figure~\ref{fig:mag_hex}(e) and the hollow state is presented in the Supplementary Information. SP-STM simulations of the two opposite magnetic domains are displayed in the insets and show a good agreement with the experiments. At first sight there is no obvious sign of the H p$(2 \times 2)$ superstructure so we neglect it for now and discuss it later.


\begin{figure}
  \centering
  \includegraphics{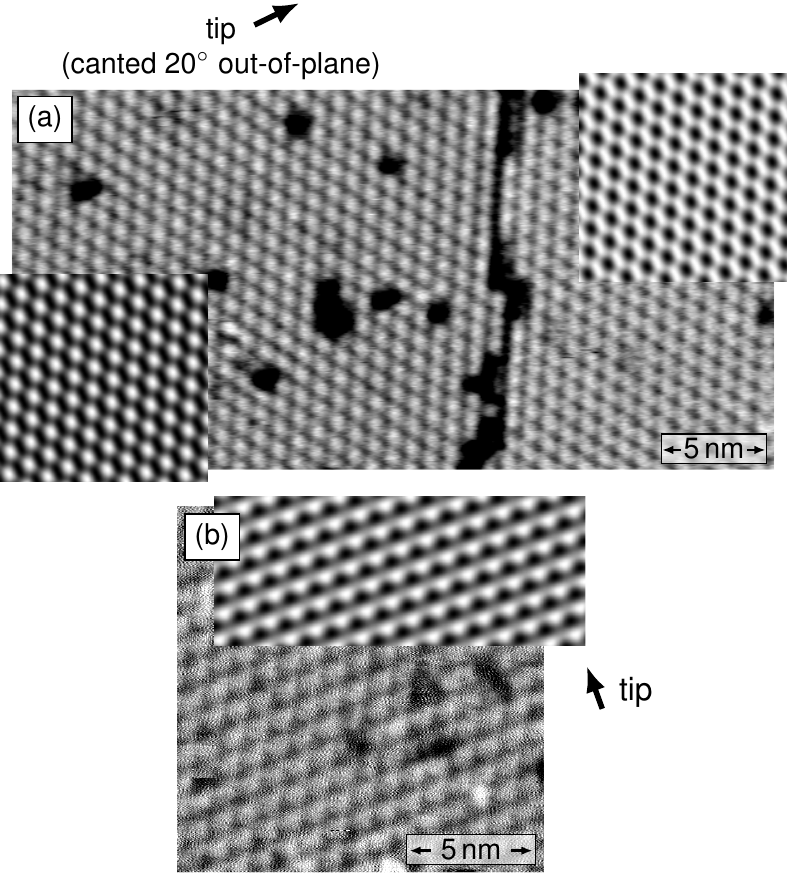}
  \caption{\textbf{In-plane components of the magnetic state in the hexagonal phase.} (a) Spin-resolved differential conductance map showing the two magnetic domains, the boundary between the domains coincides with the dark line. The image was measured with a tip sensitive slightly to the out-of-plane component of the sample magnetization and mostly to the in-plane component indicated by the arrow. The corresponding SP-STM simulations of the on-top state for both domains are shown in the insets. (b) Spin-resolved differential conductance map showing another in-plane component of the magnetic state and the corresponding SP-STM simulation of the on-top state as inset. Measurement parameters:~(a):~\SI{-700}{\milli\volt}, \SI{1}{\nano\ampere}, \SI{4}{\kelvin}, \SI{0}{\tesla}, Cr bulk tip, (b):~\SI{-1}{\volt}, \SI{1}{\nano\ampere}, \SI{8}{\kelvin}, \SI{0}{\tesla}, Fe coated W tip. The SP-STM simulations were performed with the following realistic parameters: tip-sample distance:~\SI{600}{\pico\meter}, work function:~\SI{4.8}{\electronvolt}, spin polarization of sample (s) and tip (t):~$P_sP_t=0.2$.}
  \label{fig:mag_hex_ip}
\end{figure}

In order to check that the model proposed in figure~\ref{fig:mag_hex}(e) is completely compatible with the experiments, we also performed measurements and simulations with tips sensitive to in-plane components of the sample magnetization. These results are displayed in figure~\ref{fig:mag_hex_ip}. Figure~\ref{fig:mag_hex_ip}(a) shows the two opposite magnetic domains measured with a mostly in-plane sensitive tip. The dark line corresponds to a phase domain boundary of the $ p(2\times 2)$ H superstructure and coincides with the position of the boundary between the two magnetic domains. Because the six-fold symmetry of the magnetic pattern observed with the out-of-plane magnetic tip is broken and reduced to $C_2$, we conclude that this tip has a significant in-plane magnetic component. With this tip, the magnetic state appears as a distorted checkerboard pattern, which can nicely be reproduced by the SP-STM simulations shown in the square images around the differential conductance map. Figure~\ref{fig:mag_hex_ip}(b) shows another in-plane component of the magnetic structure. In this case, the magnetic sensitivity of the tip was determined using the spin spiral in the surrounding H1 phase of the Fe double layer (not shown here). For this configuration, we see a pattern of alternating bright and dark triangles. This pattern is also found in the SP-STM simulations, which confirms the validity of our hexagonal nanoskyrmion lattice model for the spin configuration. 

\begin{figure}
  \centering
  \includegraphics{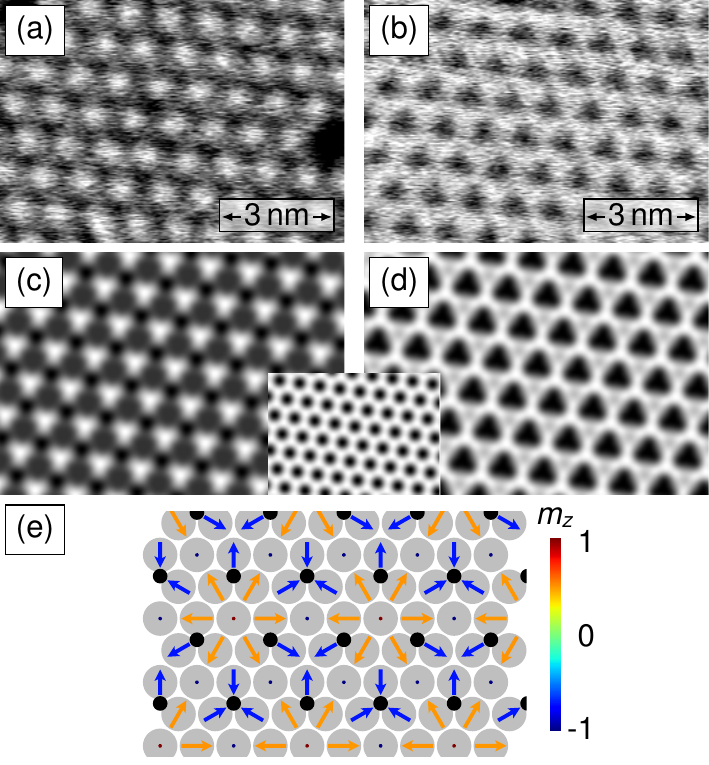}
  \caption{\textbf{Position of the H atoms in the hexagonal phase.} (a)-(b) Spin-resolved differential conductance maps showing the magnetic state in the hexagonal phase, measured with an out-of-plane sensitive magnetic tip (same data as figure~\ref{fig:mag_hex}). The images (a) and (b) show the two opposite magnetic domains. Small darker dots are visible between the bright dots in image (a) and the dark dots in image (b) have a triangular shape. These features appear because the position of the H atoms is also visible in addition to the magnetic pattern. Measurement parameters:~\SI{-1}{\volt}, \SI{1}{\nano\ampere}, \SI{4}{\kelvin}. (c)-(d) SP-STM simulations of the on-top state for an out-of-plane sensitive tip with an additional modulation reproducing the effect of the H atoms and allowing to reproduce the experimental data. The inset shows the pattern used to simulate the contribution of the H atoms. (e) Spin configuration of the on-top state showing the position of the H atoms which allows to reproduce the experimental data. The SP-STM simulations were performed with the following realistic parameters: tip-sample distance:~\SI{600}{\pico\meter}, work function:~\SI{4.8}{\electronvolt}, spin polarization of sample (s) and tip (t):~$P_sP_t=0.2$.
}
  \label{fig:H_positions}
\end{figure}

However, if we look again at the magnetic pattern in figure~\ref{fig:mag_hex} (and reproduced in figure~\ref{fig:H_positions}), some features of the measured out-of-plane data are not present in the SP-STM simulations.  Some darker dots are present between the bright dots in figure~\ref{fig:H_positions}(a) and the dark dots have a triangular shape in figure~\ref{fig:H_positions}(b), the six-fold symmetry is reduced to three-fold. An explanation is that these features do not have a magnetic origin but rather originate from the H atoms. Indeed, when the tip is only slightly sensitive to the magnetic state of the sample, the p$(2\times2)$ H superstructure appears as an hexagonal arrangement of dark dots in the differential conductance maps measured at \SI{-1}{\volt} (which is the typical voltage that we use for magnetic imaging of this system, image not shown here). This modulation of the local density of states adds up with the contrast produced by the magnetic state.  Figure~\ref{fig:H_positions} shows that by superposing such an hexagonal p$(2\times2)$ pattern with the SP-STM simulations of the magnetic state, it is possible to generate the observed additional features. The panels (a) and (b) show again the differential conductance maps of the two domains measured with an out-of-plane sensitive tip. For the domain in (a), we see dots darker than the background between the bright dots whereas in the domain shown in (b), the dark dots have a triangular shape. The simulation of the expected signal from the two domains of the magnetic on-top state, taking into account the presence of the H atoms, is presented in the panels (c) and (d). Among the eight possible positions of the magnetic unit cell with respect to the structural unit cell of the H superstructure (details can be found in the Supplementary Information), only the one illustrated by the sketch in figure~\ref{fig:H_positions}(e) allows to reproduce the experiments.

\begin{figure*}
  \centering
  \includegraphics{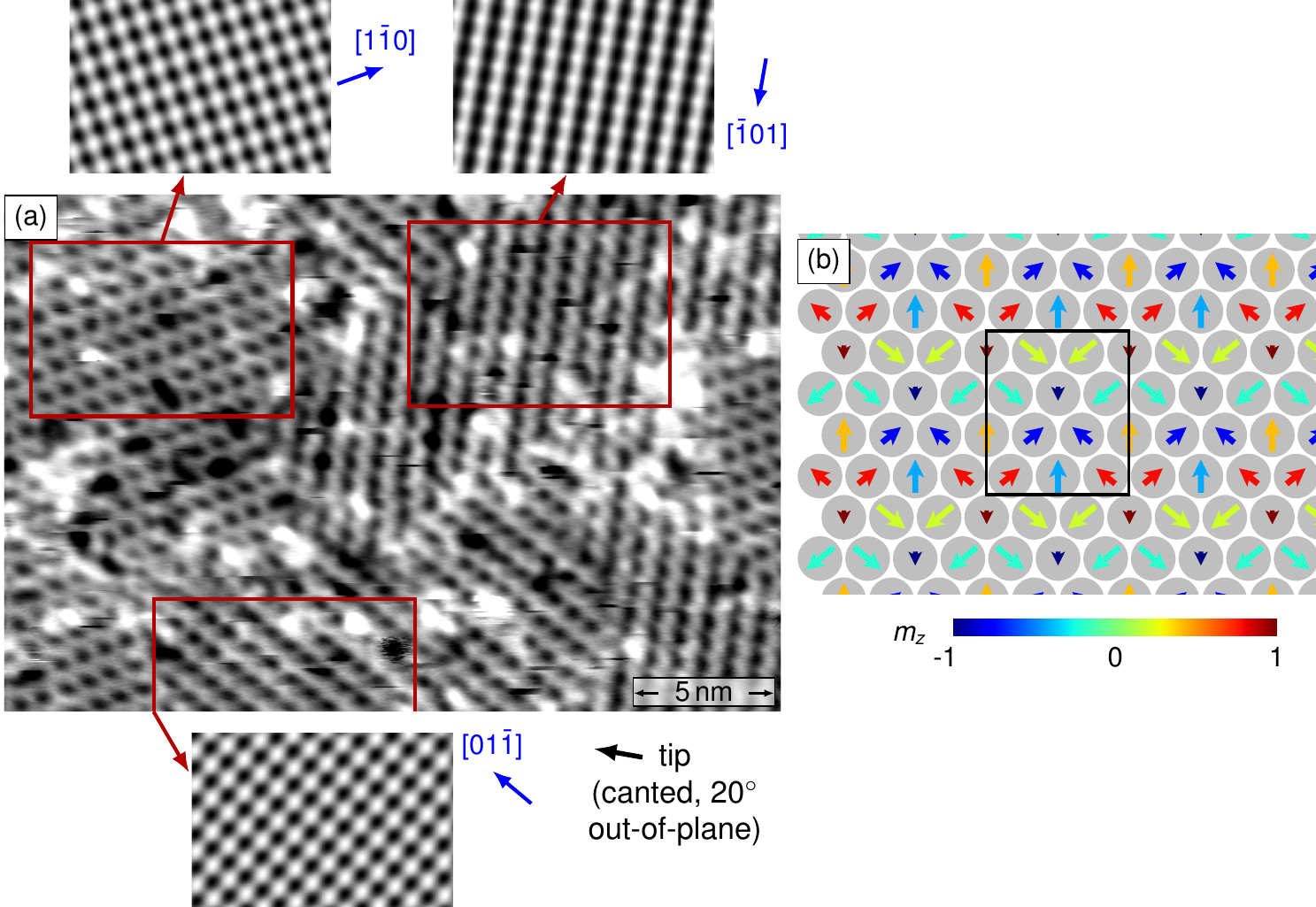}
  \caption{\textbf{Magnetic state in the square phase.} (a) Spin-resolved differential conductance map showing the magnetic state in the three rotational domains of the square H-superstructure. The three small images show SP-STM simulations corresponding to each domain for the spin configuration shown in (b). Measurement parameters:~\SI{1.4}{\volt}, \SI{3}{\nano\ampere}, \SI{4}{\kelvin}, \SI{0}{\tesla}, Cr bulk tip.  The SP-STM simulations were performed with the following realistic parameters: tip-sample distance:~\SI{600}{\pico\meter}, work function:~\SI{4.8}{\electronvolt}, spin polarization:~$P_sP_t = 0.2$.}
  \label{fig:mag_sq}
\end{figure*}

Spin-polarized measurements were also performed on the roughly square structural phase of the hydrogenated Fe monolayer. Following the symmetry of the superstructure, three rotational domains are also found for the magnetic state as illustrated in figure~\ref{fig:mag_sq}, where the three red boxes enclose areas corresponding to these three domains. We observe a rectangular pattern of magnetic origin, consisting of darker structures on a brighter background. The magnetic unit cell has twice the area of the unit cell of the H superstructure, the diagonals of the structural cell of the H superstructure correspond to the lattice vectors of the magnetic cells. The rectangular magnetic unit cell seems to be commensurate with the H superstructure (regardless of the irregularities arising from the H superstructure  with respect to the Fe lattice) and has a size about \SI{0.80}{\nano\meter} in the $[1\bar{1}0]$ direction and about \SI{0.95}{\nano\meter} in the $[11\bar{2}]$ direction. From our observations and symmetry arguments (more details are available in the Supplementary Information), we propose the model presented in figure~\ref{fig:mag_sq}(b) for the magnetic state. This state is obtained by the superposition of two cycloidal spin spirals (with normalized magnetic moments), it has a magnetic unit cell corresponding to the experimental data and exhibits a mirror symmetry. Furthermore, the out-of-plane magnetic moment is fully compensated within each cell. This is in agreement with our experiments in which opposite magnetic domains are not present and the application of an external out-of-plane magnetic field of $\pm\SI{9}{\tesla}$ does not induce any change of the magnetic configuration (not shown here).


In order to validate the model presented in figure~\ref{fig:mag_sq}(b), we also performed SP-STM simulations. Because the incorporation of a large amount of H atoms is necessary to build the roughly square superstructure, the H1 areas of the hydrogenated Fe double layer on this sample are disordered and do not exhibit any clear magnetic pattern. For this reason, they cannot be used as a reference to determine the magnetization axis of the tip for the measurement presented in figure~\ref{fig:mag_sq}(a). The tip sensitivity direction was thus inferred from the SP-STM simulations by finding a set of three magnetic domains with one tip magnetization axis that best agrees with the experimental data. The result of these simulations for the three rotational domains is displayed in the insets of figure~\ref{fig:mag_sq}(a). Our proposed magnetic model allows to reproduce the general features of the observed magnetic pattern, even though the matching is not perfect, in particular concerning the domain at the bottom of the image. These discrepancies might come from the fact that our model is a pure superposition of spirals, which is maybe not strictly the case. A distorted version of this ideal state might be more stable. In addition, similarly to the case of the hexagonal phase, the H superstructure is probably inducing a modulation of the local density of states which is added to the magnetic contrast. However, since we could not determine precisely the structural model of the roughly square hydrogenated phase, we cannot repeat the analysis performed for the hexagonal phase.

\section{Conclusion}

Our results reveal a clear coupling of the magnetic state to the H superstructures created by the H incorporation in the film. Whereas the nanoskyrmion lattice in the pristine fcc Fe monolayer on Ir(111) is incommensurate with the atomic lattice~\cite{heinze_spontaneous_2011, grenz_probing_2017, hauptmann_sensing_2017}, the magnetic unit cells in the observed states of the two hydrogenated phases are commensurate with the H superstructures. This suggests that the presence of the H atoms induces a nanoscale spatial modulation of the different magnetic energy contributions. The nanoskyrmion lattice  with respect to the Fe lattice is the result of a subtle balance between these different contributions and it appears that the incorporation of H atoms allows to modulate them and hence to determine the symmetry of the magnetic configuration. However, the strong non-collinearity of the system as well as its small typical length scale of about \SI{1}{\nano\meter} are preserved. 

Our experiments offer an additional example of the efficiency of hydrogenation to modify non-collinear magnetic states. We show that it is possible to induce changes in the complex nanoskyrmion lattice state of the atomic Fe monolayer on Ir(111) conserving its main features: strong non-collinearity and nanoscale magnetic period. We also find that the symmetry of the magnetic state depends on the amount of H atoms incorporated in the film. This offers the possibility to control easily the symmetry of the magnetic configuration in ultrathin films without the need to change the symmetry of the substrate.

\section*{Acknowledgements}
We thank A. Kubetzka for discussions and help with the experiments. Financial support by the European Union via the Horizon 2020 research and innovation programme under grant agreement No.~665095 and by the Deutsche Forschungsgemeinschaft via SFB668-A8 is gratefully acknowledged.

\end{document}